\input harvmac
\overfullrule=0pt

%

\def\e{{\epsilon}}

\def\K3{{\bf K3}}

\Title{ \vbox{\baselineskip12pt
\hbox{hep-th/9912025}
\hbox{IFP-9912-UNC}
\hbox{HUB-EP-99/62}}}
{\vbox{\centerline{Horizontal and Vertical Five-Branes}
\bigskip\centerline{in Heterotic/F-Theory Duality}}}

\centerline{Bj\"orn Andreas$^{\dagger}$\foot{bandreas@physics.unc.edu, 
supported by U.S. DOE grant DE-FG05-85ER40219} and Gottfried Curio$^*$\foot{
curio@physik.hu-berlin.de}}
\bigskip
\centerline{\it $^{\dagger}$Department of Physics and Astronomy}
\centerline{\it
University of North Carolina, Chapel Hill, NC 27599-3255, USA}
\bigskip
\centerline{\it $^*$Humboldt-Universit\"at zu Berlin,}
\centerline{\it Institut f\"ur Physik, D-10115 Berlin, Germany}\bigskip
\def\sqr#1#2{{\vbox{\hrule height.#2pt\hbox{\vrule width
.#2pt height#1pt \kern#1pt\vrule width.#2pt}\hrule height.#2pt}}}

\noindent
We consider the heterotic string on an elliptic Calabi-Yau three-fold 
with five-branes wrapping curves in the base ('horizontal' curves) of the 
Calabi-Yau as well as some elliptic fibers ('vertical' curves). 
We show that in this generalized set-up, where 
the association of the heterotic side with the $F$-theory side is changed
relative to the purely vertical situation,
the number of five-branes wrapping the elliptic fibers still 
matches the corresponding number of $F$-theory three-branes.

\Date{}

\lref\FMW{R. Friedman, J. Morgan and E. Witten, ``Vector Bundles and F- 
Theory,'' Commun. Math. Phys. {\bf 187} (1997) 679, hep-th/9701162.}

\lref\AC{B. Andreas and G. Curio, ``Three-Branes and Five-Branes in 
N=1 Dual String Pairs,'' Phys. Lett. {\bf B417} (1998) 41, hep-th/9706093.}

\lref\Raj{G. Rajesh, ``Toric Geometry and $F$-theory/Heterotic Duality in Four
Dimensions'', JHEP 12 (1998) 18, hep-th/9811240.}

\lref\RajDia{D.-E. Diaconescu and G. Rajesh, ``Geometrical Aspects 
of Fivebranes in Heterotic/F-Theory Duality in Four Dimensions'', 
hep-th/9903104.}

\lref\DOe{R. Donagi, A. Lukas, B. A. Ovrut and D. Waldram,
``Non-Perturbative Vacua and Particle Physics in M-Theory'', 
 JHEP {\bf 9905} (1999) 018, hep-th/9811168.}

\lref\DOz{R. Donagi, A. Lukas, B. A. Ovrut and D. Waldram,
``Holomorphic Vector Bundles and Non-Perturbative Vacua in M-Theory'',
hep-th/9901009.}

\lref\DOd{R. Donagi, B. A. Ovrut and D. Waldram,
``Moduli Spaces of Fivebranes on Elliptic Calabi-Yau Threefolds'',
hep-th/9904054.}

\lref\ACu{B. Andreas and G. Curio, ``On Discrete Twist and Four Flux in
N=1 Heterotic/F-Theory Compactifications'', hep-th/9908193.}

\lref\wflux{E. Witten, ``On Flux Quantization in M-Theory and the Effective
Action,'' J. Geom. Phys. {\bf 22} (1997) 1, hep-th/9609122.}

\lref\damu{K. Dasgupta and S. Mukhi, ``A Note on Low Dimensional String
Compactifications,'' Phys. Lett. {\bf B398} (1997) 285, hep-th/9612188.}

\lref\CDo{G. Curio and R. Donagi, ``Moduli in {N=1} Heterotic/F-Theory
 Duality,'' Nucl.Phys. {\bf B518} (1998) 603, hep-th/9801057.}

\lref\Berg{P. Berglund and P. Mayr, ``Stability of Vector Bundles from
F-Theory'', hep-th/9904114.}

\lref\SVW{S. Sethi, C. Vafa and E. Witten, ``Constraints on Low Dimensional 
String Compactifications,'' Nucl. Phys. {\bf B480} (1996) 213, 
hep-th/9606122.}

\lref\ACL{B. Andreas, G. Curio and D. L\"ust, ``N=1 Dual String Pairs and 
their Massless Spectra'', Nucl. Phys. {\bf B507} (1997) 175, hep-th/9705174.}

\lref\Hart{R. Hartshorne, ``Algebraic Geometry'', Springer 1977.}

\lref\KV{S. Kachru and C. Vafa, 
``Exact Results for N=2 Compactifications of Heterotic Strings'',
Nucl. Phys. {\bf B450} (1995) 69, hep-th/9505105. }

\lref\HKTY{S. Hosono, A. Klemm, S. Theisen and S.-T. Yau, 
``Mirror Symmetry, Mirror Map and Applications to Complete 
Intersection Calabi-Yau Spaces'', Nucl. Phys. {\bf B433} (1995) 501,
hep-th/9406055.}

\lref\KLRY{A. Klemm, B. Lian, S.-S. Roan and S.-T. Yau,
``Calabi-Yau fourfolds for M and F-theory compactifications'', 
Nucl. Phys. {\bf B 518} (1998) 515, hep-th/9701023.}

\lref\KMV{S. Katz, P. Mayr and C. Vafa, ``Mirror symmetry and exact solution
of 4D N=2 Gauge theories I'', hep-th/9706110.}

\lref\MNVW{J.A. Minahan, D. Nemeschansky, C. Vafa and N.P. Warner, 
``E-Strings and N=4 Topological Yang-Mills Theories'' 
Nucl. Phys. {\bf B527} (1998) 581, hep-th/9802168.}

\lref\klem{A. Klemm, private communication.}

\lref\mberg{P. Berglund and P. Mayr,''Heterotic/F-Theory Dulality from
Mirror Symmetry'', Adv.Theor.Math.Phys.2 (1999) 1307, hep-th/9811217.}



\newsec{Introduction}
The compactification of the heterotic string on a Calabi-Yau three-fold
with a vector bundle over it, which breaks part of the $E_8\times E_8$ 
gauge symmetry, leads to $N=1$ supersymmetric vacua in four dimensions.

In particular the class of elliptically fibered Calabi-Yau three-folds $Z$
has the double advantage to admit a direct construction of the 
vector bundles and 
to allow a dual description in terms of $F$-theory compactified on 
elliptically fibered Calabi-Yau four-folds $X$ \FMW. 

To obtain a consistent heterotic compactification on elliptic
Calabi-Yau three-folds, the anomaly cancellation condition requires 
to include a number $a_f$ of five-branes wrapping the 
elliptic fibers. It has been shown that these five-branes match precisely 
the number $n_3$ of space-time filling three-branes necessary for 
tadpole cancellation on the $F$-theory side \FMW,\AC\ and so giving a
heterotic string explanation of the number of three-branes.     

Here we will be interested in a slightly more general situation, where 
one admits also five-branes wrapping holomorphic curves  
in the base $B_2$ of the elliptic Calabi-Yau three-fold
(we will take $B_2$ to be
a Hirzebruch surface $F_n$ of $n=0,1,2$ or a del Pezzo surface $dP_k$, 
$k=0,\dots ,8$; the case of an Enriques surface could be treated along
similar lines). 
The picture which emerges is, the five-branes wrapping the elliptic 
fibers map to the $F$-theory three-branes whereas a five-brane 
wrapping a curve in the base $B_2$ of the elliptic Calabi-Yau three-fold $Z$ 
corresponds to a blow-up of the base $B_3$ along the curve in the common 
$B_2$ of the elliptic Calabi-Yau four-fold $X$ on the $F$-theory side 
(recall that $B_3$ resp. $X^4$ is a $P^1$ resp. $K3$ fibration over $B_2$).

Several aspects of this generalized set-up were discussed in 
\mberg,\Raj,\RajDia, mainly in the case of wrapping curves of genus zero in 
the base and in the extreme situation when the bundles have trivial structure 
group leading to an unbroken $E_8 \times E_8$ gauge group. 
The possibility of wrapping five-branes on curves in the base (or even in $Z$)
was also carefully discussed with many examples in \DOe,\DOz,\DOd. 
Note that the $g$ $U(1)$ vector multiplets ($g$ being the genus of $C$) 
coming from reduction of the two-form of self-dual field-strength on the
five-brane, are associated with corresponding modes from three-cycles
in the blown-up base $\tilde{B}_3$ and the four-form of self-dual 
field-strength of type IIB on the $F$-theory side. 

Note that a curve lying in the base $B$ will have in general an interesting
set of deformations which will play a role below.
In the set-up with five-branes wrapping only fibers
the brane-parameters on both sides took care of themselves. 
Remember that the 'three coordinates' of such a three-brane, considered
as a point in the ${\bf P}^1$ bundle $B_3$ over the visible $B_2$, are
simply given by the base-point of the corresponding elliptic fiber on the
heterotic side which is trivially given on the common $B_2$ and the
'third coordinate', i.e. the fact that there is still a moduli space
${\bf P}^1$ in the game,
is given by a complex scalar whose real resp. imaginary part is given by 
the position in the $M$-theory interval $S^1/Z_2$ in the eleventh dimension 
resp. the axion $a$ corresponding to the reduction of the self-dual three-form
on the five-brane (note that this $S^1$ must collapse to a point at the 
boundaries of the interval - thus leading to the third, missing complex
coordinate, parametrizing on the $F$-theory side the $P^1$ fiber over the
common $B_2$ - as there one has a transition to $E_8$ instantons without
the interval position modulus and so without $a$ too (the five-brane
three-form being odd it must vanish at the orbifold fixed planes anyway); 
cf. \RajDia, \DOd ).
Note also that the heterotic string
on a wrapped $T^2$ is equivalent to the $SO(32)$ heterotic string and a
small instanton-fivebrane has a non-perturbative $SU(2)$ which has
as moduli space on $T^2$ exactly a ${\bf P}^1$ (in general
an $SU(n)$ bundle on $T^2$ has as moduli space a ${\bf P}^{n-1}$).

Of course the expectation is still that the five-branes wrapping 
elliptic fibers correspond to the $F$-theory three-branes and this should not
be changed by the occurrence of the 'other', horizontal five-branes. However
the actual proof of the matching in the purely vertical case \FMW\ did not
associate these objects directly with each other (although of course this
is the obvious underlying intuition) but rather computed the number of the
relevant five-branes resp. three-branes independently on both sides in data
of the common base $B_2$. To get the matching one had to use then 
an association of both sides
(this was the relation $\eta_{1,2}=6c_1(B_2)\pm t$, respecting the condition
$\eta_1+\eta_2=12c_1(B_2)$, where $t$ characterizes the $P^1$ fibration of 
$B_3$ over $B_2$). But we will have a non-trivial
horizontal five-brane (of class $W_B$) and so 
the mentioned condition changes to 
$\eta_1+\eta_2+W_B=12c_1(B_2)$ on the one hand and $B_3$ is modified by the
blow-up on the other hand. {\it This makes the numerical matching non-trivial.}

In order to proof the matching of vertical five-branes and three-branes 
in this generalized set-up we will facilitate the analysis by considering a 
five-brane wrapping a smooth irreducible curve $C$ 
representing a class $W_B$ and thus blow up once the
$F$-theory base $B_3$. We will separate our analysis into parts starting with 
the simplest case of a heterotic string compactification on $Z$ with an 
$E_8\times E_8$ vector bundle corresponding under duality to $F$-theory
on a smooth $X$ (where smooth means that the elliptic fibers of $X$ 
degenerate in codimension one 
not worse than $I_1$ in the Kodaira classification). After 
explaining how a blow-up of $W_B$ in the base $B_3$ of $X$ changes the 
Euler number of $X$ and 
therefore the number of three-branes, we follow the strategy of 
\FMW\ and express the number of three-branes and five-branes in 
comparable data on the common base $B_2$.
Then we proceed to models with an {\it A-D-E} gauge group
and consider a heterotic model with an $E_8\times V_2$ 
vector bundle where $V_2$ can be $E_7, E_6, SU(n)$ giving 
corresponding unbroken gauge groups. The dual $F$-theory model 
(assuming no further monodromy effects, which could alter the gauge group,
being present there)
is described by compactifying on an $X$ whose elliptic 
fibers degenerate of type {\it A-D-E} . For simplicity we will admit only 
that the fibers degenerate over codimension one in $B_3$ (which
can be done by adjusting the ${\bf P^1}$ bundle over $B_2$ respectively the 
$\eta$ class on the heterotic side). In order to 'fill out'  
$V_2$ with enough instanton number (measured by the $\eta$-bound) we have 
to perform the $W_B$-{\it shift} in the $E_8$ bundle. 
One has, just as in the smooth case, again an explicit formula for the
Euler number, which was recently proved in \ACu. However, as this ansatz is not
immediately adapted to the blow-up procedure which destroys the usual
$P^1$ fibration structure of $B_3$, we will proceed in the proof of the
five-brane/three-brane correspondence more easily by adopting the 
strategy of \AC.
In the latter one expresses the number of three-branes again via 
the Euler number, which this time is then  in turn expressed 
by the Hodge numbers, they again by the heterotic data (including the 
index of bundle moduli 'inside' the complex structure deformations of $X$);
then the close similarity of the expression for the index of bundle moduli 
with the expression for the second Chern class of the vector bundle is used
to transport directly the number of five-branes to the $F$-theory side.
We will reduce our more general situation ($W_B\neq 0$) to this original
case ($W_B=0$) by collecting the changes in the Hodge numbers of the fourfold
induced by the blow-up procedure resp. by the 'change' in heterotic bundle
moduli (in a sense made precise below) when transported to the $F$-theory side;
taken together with the deformations of the chosen curve $C$ these changes
will then cancel out in total.

In {\it section 2} we outline the general procedure of the $\eta$-shift
induced by the occurrence of a non-trivial $W_B$.
 
In {\it section 3} we show the three-brane/five-brane correspondence for 
the case of a smooth fourfold.

In {\it section 4} we lay the ground for the treatment of the (codimension
one) singular case. We recall some points of the identification of the 
moduli spaces and the expression of the Hodge numbers of the fourfold
in heterotic data; starting from there we show how this already gave
the result $a_f=n_3$ in the simple case $W_B=0$ (\FMW\ showed this in the
{\it smooth} case by a different procedure; we build on this type of argument
in section 3). Then we derive the ({\it functional}, this will be explained 
below) change in the index of bundle moduli under the shift in $\eta$ by $W_B$.

In {\it section 5} the necessary informations about deformations of $C$ are
derived. For this we investigate the corresponding question for $C$ in the 
base surface $B=B_2$ and in the elliptic surface ${\cal E}$ which lies in $Z$
above $C$.

In {\it section 6} we put our informations about ({\it functional}) 
changes in the Hodge numbers of the fourfold together and show that they
cancel each other thereby reducing the proof of our result to the original
case ($W_B=0$).


\newsec{Testing $a_f=n_3$ with $W_B$ turned on}

We consider the heterotic string on $\pi:Z\rightarrow B_2$ with a section 
$\sigma$ and specify a vector 
bundle $V=E_8\times V_2$ where we fix $V_1$ to be an $E_8$ bundle and 
$V_2=E_8, E_7, E_6$ or $SU(n)$.

Under the decomposition $\sigma H^2(B_2)+H^4(B_2)$, 
where the base cohomology is pulled back by $\pi^*$,
the second Chern class of $V_2$ respectively the fundamental 
characteristic class of $E_{8,7,6}$ bundles, is given by 
\eqn\v{\lambda(V_2)=\eta\sigma+\pi^*(\omega)}
where $\lambda(V_2)=c_2(V_2)$ for $V_2=SU(n)$ and $\lambda(V_2)=c_2(V_2)/C$ 
with $C=60,36,24$ for $E_8, E_7, E_6$ bundles. Note further that 
$\eta\in H^2(B_2)$ is {\it arbitrary} and $\omega\in H^4(B_2)$ is 
determined in terms of $\eta$. The explicit expressions for the 
characteristic classes of $E_8$ bundles and $SU(n)$ bundles are given 
in \FMW\ for $E_7$ and $E_6$ bundles see \ACu. 
A corresponding decomposition of the second 
Chern class of $Z$ gives
\eqn\vc{c_2(Z)=12c_1\sigma+c_2+11c_1^2 .}
The general condition for anomaly cancellation with five-branes is 
\eqn\ws{\lambda(V_1)+\lambda(V_2)+W=c_2(Z)}
where $W$ is the cohomology class of the five-branes. As we want to treat the
case that one has, besides some elliptic fibers wrapped by five-branes
(the case already treated in \FMW ), also a wrapped curve $C$ in the base 
let us now assume
for the five-brane class a corresponding decomposition in a part consisting of
base cohomology (to be considered embedded via $\sigma$) and a fiber part
\eqn\vx{W=W_B+a_fF}
where $W$ has to be effective \DOe, which in the cases considered
is equivalent to $W_B$ being effective and $a_f\geq 0$ \DOz .
Now the $\sigma H^2(B_2)$ part of \vx\ gives the condition
\eqn\vxb{\eta_1+\eta_2+W_B=12c_1}
On the other hand one gets an
expression for the number of five-branes wrapping still just the 
elliptic fiber of $Z$
\eqn\qw{a_f=12+10c_1^2-\omega_1-\omega_2 .} 
This gives also a prediction for the number of three-branes on the 
$F$-theory side which is proportional to the Euler number of the 
Calabi-Yau four-fold, one has 
$\chi(X)/24=n_3$. Now the fibration structure of the $F$-theory base $B_3$ 
is described by assuming the ${\bf P^1}$ bundle over $B_2$ to be a 
projectivization
of a vector bundle $Y={\cal O}\oplus {\cal T}$, where ${\cal T}$ is 
a line bundle 
over $B_2$ and the cohomology class $t=c_1({\cal T})$ encodes the ${\bf P^1}$ 
fibration structure. Now the duality with the heterotic side is 
in the case $W_B=0$ implemented 
by choosing the $E_8\times V_2$ bundle so that the $\eta$ class of the 
$E_8$ bundle is given by $\eta_1=6c_1(B_2)+t$ and the $\eta$ class of the 
$V_2$ bundle by $\eta_2=6c_1(B_2)-t$. 

As mentioned, {\it we will now allow five-branes to wrap a curve $C$ of class
$W_B$ in the base $B_2$. This modifies the direct relation between
the $\eta$ classes $\eta_{1,2}=6c_1\pm t$ 
and the ${\bf P}^1$ fibration described by $t$}.
We will describe what makes it possible to have still the relation
$a_f=n_3$. In accordance with condition \vxb we must deviate
from the $\eta_{1,2}=6c_1\pm t$ set-up by redefining 
\eqn\shif{\eta_1\rightarrow \eta_1-W_B .}
Before we come to the actual computations let us note that in
general the relation for the number of three-branes will be modified 
due to the appearance of the $F$-theory analog of $M$-theory four-flux
which one is even forced to turn on if $\chi(X)/24$ is 
not integral \wflux,\damu. One has the generalized relation  
\eqn\gfl{{\chi(X)\over 24}=n_3+{1\over 2}\int_X G\wedge G.}
In particular for $SU(n)$ bundles it was expected from an 
argument given in \CDo\ that $G^2=-\sum_{i}{1\over 2}\pi_*(\gamma_i^2)$, 
i.e there
is a relation between four-flux and discrete twist (which appears in the 
$SU(n)$ bundle construction). Such a relation could be proved indirectly 
by an explicit computation in \ACu. 

Since a vector bundle is determined by specifying the $\eta$ class
which has to satisfy a bound \mberg,\Raj,\ACu,\Berg\ let us note here that 
(the form of) the bound of \Raj\ (for the {\it new} $\eta$) 
is not modified if we include a non-zero $W_B$ as discussed in \ACu.  

\newsec{The smooth case}

As mentioned in the introduction, we consider 
here the heterotic 
string on $Z$ with an $E_8\times E_8$ bundle corresponding to $F$-theory
on a smooth $X$. In order to perform the $\eta$-shift in the vector bundle,  
let us recall that the fundamental characteristic class of 
an $E_8$ bundle is 
\eqn\b{\lambda(V)=\eta\sigma-15\eta^2+135\eta c_1-310c_1^2}
and the anomaly cancellation condition then determines the number of 
five-branes $a_f$ wrapping elliptic fibers of $Z$. Now we have two 
choices to perform the {\it shift} either in the first or in the second 
$E_8$ factor. For the shift in $\eta_1$ we get  
\eqn\c{a_f=c_2+91c_1^2+30t^2+15W_B^2-45W_Bc_1-30W_Bt}
whereas for shifting $\eta_2$ we get
\eqn\cc{a_f=c_2+91c_1^2+30t^2+15W_B^2-45W_Bc_1+30W_Bt}
which for vanishing $W_B$ of course reduces to the known expression
derived in \FMW. Note that if we perform the shift (assuming here $t\ge 0$)
in $\eta_1$
the $\eta$-bound requires $W_B\le c_1+t$ whereas for $\eta_2$ the bound
requires $t\le c_1$. Now if we perform the shift in $\eta_2$ then the \
bound for that bundle requires $t+W_B\le c_1$ whereas from $\eta_1$ comes
no further condition to $t$ since $6c_1+t>5c_1$.    

On the other hand, the number of three-branes $n_3=\chi(X)/24$ to be included
for a consistent $F$-theory compactification on a smooth $X$, as computed 
in \SVW, is
\eqn\ccc{n_3=12+15\int_{B_3}c_1(B_3)^3.}
We will be interested now to see how the effect of wrapping a five-brane
on a curve $C$ in $B_2$ of class $W_B$ on the heterotic side, reflected
in a blow-up along $C$ in $B_3$, will change the number of three-branes on
the $F$-theory side. A blow-up of $C$ in $B_3$ produces a three-fold
$\pi: {\tilde B_3}\rightarrow B_3$ with 
\eqn\d{c_1(\tilde B_3)=\pi^* c_1(B_3)-E}
where $E$ denotes the exceptional divisor, a ruled surface over $C$. 
It is useful to recall here that the well known case of a 
blow-up of a point on a surface
leading to an exceptional ${\bf P^1}$ of self-intersection number $-1$
generalizes to a relation (in $H^4(\tilde{B_3})$) 
for the ruled surface over $C$ ($E^3$ is here a number which occurs 
as prefactor of $l$)
\eqn\e{E^2=-\pi^*W_B-E^3 \, l}
where $l$ denotes the fiber of the ruled surface $E$ with $E\cdot l=-1$ (for a 
proof of this relation cf. \ACu). With this in 
hand we can proceed and compute the number of three-branes which are expected
to match the $a_f$ five-branes. What we find is
\eqn\g{n_3=12+15\int_{B_3}c_1(B_3)^3-3W_Bc_1(B_3)-E^3.}
Since $B_3$ is a ${\bf P^1}$ bundle over $B_2$ obtained by projectivizing
a vector bundle, as explained above, the adjunction formula gives 
\eqn\adj{c_1(B_3)=c_1+2r+t.}
The triple intersection of the ruled surface $E$ is given by (cf. \ACu )
\eqn\f{E^3=-\int_{W_B}c_1(N_{B_3}C)=-(c_1(B_3)W_B-\chi(C))}
With $12=\int_{B_2}c_1^2+c_2$ for the rational $B_2$
and that $r(r+t)=0$ in the cohomology ring of $B_3$ (i.e. the sections
of the line bundles ${\cal O}(1)$ and ${\cal O}(1)\otimes {\cal T}$ over $B_2$
cf. \FMW\ have no common zeros) and noting that $W_B$ is embedded via $W_Br$ 
into $B_3$ so that the term $W_Br$ is actually $W_Br^2$, leading after 
integration over the ${\bf P^1}$ fibers of $B_3$ to $-W_Bt$, one gets with 
adjunction $\chi(C)=-W_B(W_B-c_1)$ 
the final expression for the number of three-branes expressed in terms 
of $B_2$ data
\eqn\h{n_3=\int_{B_2}c_2+91c_1^2+30t^2+15W_B^2-45W_Bc_1+30W_Bt}
matching the $a_f$ heterotic five-branes in case of the shifted $\eta_2$ 
(for the $\eta_1$ shift one has to replace $t$ by $-t$ on the $F$-theory side).

\newsec{The singular case I: comparison of moduli spaces}

We will consider here the situation where the fibers of the Calabi-Yau
four-fold degenerate of type {\it A-D-E} over codimension one 
(actually over $B_2$) in $B_3$
corresponding to an unbroken gauge group of the same type which on the 
heterotic side comes from an $E_8\times V_2$ bundle over $Z$.

The codimension one degeneration is established (cf. \ACu) for the 
{\bf A} series by setting $t=c_1$, for the {\bf D} series only in 
the case $D_4$ a codimension one condition can be established for 
$t=2c_1$ and for the {\bf E} series the conditions are 
$t=3c_1,4c_1$ for $E_6$ resp. $E_7$.  

Note that we have to perform 
the {\it shift} always in the $E_8$ bundle since the choice of 
$t$ (for codimension one) sets the $\eta$ class on its lower bound. 

Now, in the case of a codimension one degeneration of the elliptic fiber
one can write down a general expression for the Euler number of the 
corresponding Calabi-Yau four-fold (which was first written down 
in \HKTY\ based there on toric computer analysis and proved in \ACu ) and 
is given by 
\eqn\Hkty{\chi(X)=288+360\int_{B_3}c_1(B_3)^3-r(G)c(G)(c(G)+1)
\int_{B_2}c_1(B_2)^2}
where $r(G)$ and $c(G)$ denoting the rank and the Coxeter number of the 
{\it A-D-E} gauge group $G$ respectively. If however now one 
would like to apply the blow-up procedure on the basis of this formula 
in order to get the change in the number of three-branes, 
this could not be done directly 
since the fibration structure of $B_3$ changes under the blow-up.
Although one could refine the derivation of \Hkty\ in the following 
we will discuss (as considered in the introduction) 
a different approach to test $a_f=n_3$ which will provide us also
with the change in the hodge numbers after the blow-up. 

The shift redefinition described above for $\eta_1$
will change of course $a_f$ and so better $n_3$ too. But this
is not the point of what we will follow, as {\it these} changes
are still captured by the transport provided by the re-expression (cf. below)
between the number of bundle moduli and the second Chern class of $V_2$. 
What we will follow is {\it how this transport
changes}. This is a change caused just by the change in the 
$\sigma H^2(B_2)$ part of $c_2(V_2)$, whereas the changes which occurs
because of just using 'another' $\eta_1$ occurs in the $H^4(B_2)$
part (whose changing influence on $a_f$ is still captured by the 
original construction).

Therefore let us first recall some facts about the general
comparison of the $F$-theory and heterotic moduli spaces and spectra. These 
will provide us with relations expressing the Calabi-Yau four-fold 
hodge numbers in terms of heterotic data. We will then show 
how these relations have to be modified when on the heterotic side 
five-branes wrapping curves in the base are included.

The moduli in a 4D N=1 heterotic compactification on an elliptic CY, 
as well as in the dual $F$-theoretic compactification, break into "base" 
parameters which are even (under the natural involution of the elliptic 
curves), and "fiber" or twisting parameters; the latter include a 
continuous part which is odd, as well as a discrete part. 
In \CDo\ all the heterotic moduli were interpreted
in terms of cohomology groups of the spectral covers, 
and identified with the corresponding $F$-theoretic moduli in a certain 
stable degeneration. For this one uses the close connection of 
the spectral cover and the {\it A-D-E} del Pezzo fibrations. 
For the continuous part of the twisting moduli, this amounts 
to an isomorphism between certain abelian varieties: the connected 
component of the heterotic Prym variety (a modified Jacobian) and the 
$F$-theoretic intermediate Jacobian. The comparison of the discrete part,
involving gamma class and four-flux, 
refines the matching of  $a_f$ five-branes and $n_3$ three-branes, as
mentioned above.

By working with elliptically fibered $Z$
one can extend adiabatically the known results about moduli spaces of 
$G$-bundles over an elliptic curve $E=T^2$, of course taking into account
that such a fiberwise description of the isomorphism class of a bundle 
leaves room for {\it twisting along the base $B_2$}. The latter
possibility actually involves a two-fold complication: there is a 
continuous as well as a discrete part of these data. It is quite easy to
see this  for  $G=SU(n)$:  in this case $V_2$ can
be constructed via push-forward of the 
Poincare bundle on the spectral cover $C \times _BZ$, possibly twisted
by a line bundle ${\cal N}$ over the spectral surface $C$ (an $n$-fold
cover of $B_2$ (via $\pi$) lying in $Z$), whose first Chern class 
(projected to $B_2$)
is known from the condition $c_1(V_2)=0$. So ${\cal N}$ itself is known
up to the following two remaining degrees of freedom: 
first a class in $H^{1,1}(C)$ which projects 
to zero in $B$ (the discrete part), and second an element of 
$Jac(C):=Pic_0(C)$ (the continuous part; the moduli odd under the 
elliptic involution).

The continuous part is expected
\FMW to correspond on the $F$-theory side to the odd moduli,
related there to the intermediate Jacobian 
$J^3(X)=H^3(X,{\bf R})/H^3(X,{\bf Z})$ of dimension 
$h^{2,1}(X)$, so that the following picture emerges 
(ignoring the Kahler classes on both sides). The moduli space 
${\cal M}$ of the bundles is fibered ${\cal M}\rightarrow {\cal Y}$, with
fiber $Jac(C)$. There is a corresponding  picture on the 
$F$-theory side:  the moduli 
space there is again fibered. The base is the moduli space of
complex deformations,
the fiber is (up to the discrete $G$-flux twisting parameters)
the intermediate Jacobian, one has
\eqn\hzze{h^{2,1}(Z)+h^1(Z,ad \, V)+1=h^{3,1}(X)+h^{2,1}(X).}
Here the number of bundle moduli $h^1(Z,ad \, V)=n_e+n_o$, 
even or odd
under the involution coming from the involution on the elliptic fiber, 
can be computed using a fixed point theorem \FMW\ and 
then first effectively computing the character-valued index $I=n_e-n_o$. 
So one gets
\eqn\nodd{h^1(Z,ad \, V)=I+2n_o}
where the index $I$ is given by an integral over the 
fixed point set \FMW\ and can be expressed in terms of the 
characteristic class of $V$ (cf. the last section and \ACL) ($rk=r(V)$)
\eqn\ind{I=rk-4(\lambda(V)-\eta\sigma)+\eta c_1.}
Note that this expression applies to vector bundles which are 
invariant under the involution
of the elliptic fiber ($\tau$-invariant) which is 
the case for $E_8$, $E_7$ and $E_6$ bundles (whose characteristic classes
$\lambda(V)$ were computed using the 
parabolic bundle construction which includes no 
additional twist which would break the $\tau$-invariance); also 
$SU(n)$ bundles 
where $n$ is even are $\tau$-invariant but bundles with  
$n$ odd (since one can twist with a line bundle in the spectral cover bundle
construction introducing thereby an additional term into $\lambda(V)$ ) 
are in general not $\tau$-invariant, however, the codimension
one condition on the $F$-theory side (specifying $t$) leads to the vanishing 
of the additional term in $\lambda(V)$ and in this case bundles with $n$ odd
are $\tau$-invariant. 

Note also that one can give \CDo\ an interpretation
of all the bundle moduli in terms of even respectively 
odd cohomology of the spectral surface, including an 
interpretation of the index as giving essentially the 
holomorphic Euler characteristic of the spectral surface.
More precisely one can identify the number of local complex 
deformations $h^{2,0}(C)$ of $C$ with $n_e$
respectively the dimension $h^{1,0}(C)$ of $Jac(C):= Pic_0(C)$ with $n_o$.

In this way one gets from a spectrum comparison 
the following relations \AC,\ACL\   
(where we assume that the elliptic fiber degenerates over codimension 
one in $B_3$)
\eqn\hdg{\eqalign{
h^{1,1}(X)&= h^{1,1}(Z)+1+r \cr
h^{2,1}(X)&= n_o            \cr 
h^{3,1}(X)&= h^{2,1}(Z)+I+n_o+1}}
{}From these relations one finds (this is recalled briefly below) 
that $a_f=n_3$ in the case that all five-branes wrap elliptic fibers only. 
For this one first expresses (from the heterotic identification) 
the Hodge numbers of $X$ purely in data of the common base $B_2$, 
then one uses the expression for the index $I$ (cf. \AC ) and finally 
takes into account that on the $F$-theory side the $h^{2,1}(X)$ classes 
correspond to modes odd under the $\tau$ involution on the heterotic side.

For this recall that for an $SU(n)$ bundle one has \FMW (cf. below for
a discussion of the influence of the discrete twisting parameter)
\eqn\secocher{\eqalign{c_2(V)&=\eta\sigma-{n^3-n\over 24}c_1(B_2)^2
-{n\over 8}\eta (\eta-nc_1(B_2)) \cr
 &=\eta\sigma +\omega }}
Furthermore the index $I=n_e-n_o$ is computed as \FMW, \CDo (cf. also the 
appendix of \ACu )
\eqn\indecform{I=n-1+{n^3-n\over 6}c_1(B_2)^2+{n\over 2}\eta (\eta-nc_1(B_2))
+\eta c_1(B_2)}
so that $I=I_1+I_2$ becomes with $rk=rk_1+rk_2$, the sum of the ranks of the
two vector bundles (so that $r=8+8-rk$ is the rank of the unbroken gauge group)
\eqn\indexsum{I=rk-4(\omega_1+\omega_2)+(\eta_1+\eta_2)c_1}
leading with \vxb\ to the result indicated above. Note that the index for 
$SU(n)$ bundles was computed in \FMW\ and for $E_8$ bundles in \ACL.

Before we proceed to the refinements in the situation $W_B\neq 0$ let us 
briefly recall how one proceeds from in the argument for $n_3=a_f$ {\it in 
the case} $W_B=0$ \AC. From the last expression for the total index one gets
\eqn\indexexpr{\eqalign{
I&=rk-4(c_2(V_1)+c_2(V_2))+48c_1(B_2)\sigma +12c_1^2(B_2) \cr
 &=rk-4(12c_1(B_2)\sigma +c_2(B_2)+11c_1^2(B_2))
+4a_f+48c_1(B_2)\sigma +12c_1^2(B_2) \cr
 &=rk-(48+28c_1^2(B_2))+4a_f}}
{}From $\chi(X)/6 -8=h^{1,1}(X)-h^{2,1}(X)+h^{3,1}(X)$ (cf. \SVW ), i.e.
\eqn\threebr{n_3=2+{1\over 4}(h^{1,1}(X)-h^{2,1}(X)+h^{3,1}(X))}
and the relations
\eqn\hodgehelprel{\eqalign{
h^{1,1}(X)&= 12-c_1^2(B_2)+r \cr
h^{3,1}(X)&= 12+29c_1^2(B_2)+I+n_o}}
which are refinements of \hdg\ using \foot{we assume that there is only one 
cohomologically independent section $\sigma$ of the elliptic fibration
$\pi: Z\rightarrow B$}
$h^{1,1}(Z)=h^{1,1}(B_2)+1=c_2(B_2)-1=11-c_1(B_2)^2$ 
and $\chi(Z)=-60 c_1^2(B_2)$ (cf. \KLRY ) then one 
computes indeed\foot{Note that 
in principle one has in $\omega=\omega_{\gamma =0}-{1\over 2}\pi_* \gamma^2$
a $\gamma$ related term which does, in contrast to $-4\omega$, not occur 
in the expression for the index \indecform\ 
(here $\gamma$ is the discrete twisting parameter, cf. \FMW,\ACu,\CDo, 
which in some cases even has to be present); 
therefore this $\gamma$ related 
term will also appear besides $a_f$ in \indexexpr ; nevertheless we have 
suppressed this term, for we will assume, as motivated in \CDo\ and indirectly 
proved for the codimension one singular case in \ACu, that it corresponds
with a four-flux term on the $F$-theory side, so that by \gfl\ the
sought for direct relation between $a_f$ and $n_3$ is still maintained.}
\eqn\oldres{n_3=2+{1\over 4}(12-c_1^2(B_2)+16-rk+12+29c_1^2(B_2)+I)=a_f}

So far the original argument. Now, {\it in the case} $W_B\neq 0$
observe that after the redefinition shift has been made, the 
index $I=I_1+I_2$ can be rewritten as
\eqn\inex{I=I_{old}-W_Bc_1=rk-4(\omega_1+\omega_2)+12c_1^2-W_Bc_1}
which is the usual index (of course depending now on the {\it new} $\eta$)
shifted by $\Delta I=-W_Bc_1$.

In the general case one has besides the geometric and bundle moduli
given above also to take into account 
the possible deformations $def_Z C$ of the 
actually chosen curve $C$ (which is wrapped by the five-brane) inside 
the cohomology class $W_B=[C]$ which we are going to describe now.     

\newsec{The singular case II: deformations of $C$}

We will treat first the deformations of $C$ in $B$ (always considered to
be embedded via $\sigma$ in $Z$) and in the elliptic surface
${\cal E}$ lying above $C$.

Note that in the following we will in the case of possible doubt
where, like for $C$, a self-intersection is to be taken,
denote the self-intersection of $b$ considered as a curve in the
surface ${\cal E}$ by $C^2_{{\cal E}}$ to distinguish it from the
self-intersection $C^2_{B_2}$ in $B=B_2$

\subsec{The deformations of $C$ in $B$}

We have a 'local' information $h^0(C,N_B C)$ about deformations of $C$ in $B$
as well as a 'global' one ${\rm def}_{B_2}(C)=h^0(B_2,{\cal O}(C))-1$.
Using Riemann-Roch
\eqn\ro{\sum_{i=0}^{2}(-1)^ih^i(B_2,{\cal O}(C))=
h^0(C,N_{B_2}C)-h^1(C,N_{B_2}C)+\chi (B_2,{\cal O})}
we get (with $\chi (B_2,{\cal O})=1$ from Noether for our rational $B_2$)
\eqn\defs{{\rm def}_{B}(C)=h^0(C,N_{B_2}C)-h^1(C,N_{B}C)+s-
h^2(B,{\cal O}(C))}
with the local terms
\eqn\deor{h^0(C,N_B C)-h^1(C,N_B C)={1\over 2}\chi(C)+
{\rm deg} N_B C={{Cc_1+C^2}\over{2}}}
Let us investigate the two higher cohomological corrections.
First for any curve $C$ on a rational surface $B$ 
\eqn\htu{h^2({\cal O}_B(C))=h^0({\cal O}(K-C))=0}
which can be seen from the exact sequence
\eqn\esq{0\rightarrow {\cal O}_B\rightarrow {\cal O}_B(C)\rightarrow 
{\cal O}_C(C)\rightarrow 0}
whose associated long exact cohomology sequence reads
\eqn\lesq{\eqalign{
0& \rightarrow h^0(B,{\cal O}_B)\rightarrow h^0(B,{\cal O}_B(C))
\rightarrow h^0({\cal O}_C(C)) \cr
 & \rightarrow h^1(B,{\cal O}_B)\rightarrow h^1(B,{\cal O}_B(C))
\rightarrow h^1({\cal O}_C(C)) \cr
 & \rightarrow h^2(B,{\cal O}_B)\rightarrow h^2(B,{\cal O}_B(C))
\rightarrow h^2({\cal O}_C(C))\rightarrow \cdots}} 
Here vanishing of the last term shown and $p_g(B)=0$ show together that 
$h^2({\cal O}_B(C))=0$.

Now we have to consider the {\it superabundance} $s=h^1(B,{\cal O}(C))$ 
for the cases of $B=B_2$ being a del Pezzo surface 
$dP_k$ (${\bf P}^2$ blown-up in $k$ points) or a Hirzebruch surface 
$F_n$ (a ${\bf P}^1$ bundle over ${\bf P}^1$). 
The long exact sequence shows with $q(B)=p_g(B)=0$ that 
$h^1(B,{\cal O}_B(C))=h^1({\cal O}_C(C))=h^0(C,K_C-N_{C/B})$ 
which vanishes certainly for $0>{\rm deg}( K_C-N_{C/B})=C(K_B+C)-C^2=CK$
which is guaranteed if we assume $-K$ ample;
thus for $B_2=F_n$ with $n=0,1$ and $dP_k$ with $k\neq 9$ the superabundance
vanishes. In general we have the important consequence
\eqn\scond{C\cdot c_1(B_2)>0 \Rightarrow s=0}
so that we find under this assumption
\eqn\resBdef{C\cdot c_1(B_2)>0 \Rightarrow {\rm def}_{B}(C)=h^0(C,N_{B_2}C)= 
{{Cc_1+C^2}\over{2}}.}
Let us now discuss the case $B_2=F_2$. Recall that for $F_n$
the Kaehler cone (the very ample classes) equals the positive 
(ample) classes and is given \Hart\ by the 
numerically effective classes $xb+yf=(x,y)$ 
with $x>0, y>nx$ where $b=b_-$ is the
section with $b^2=-n$ and $f$ the fiber. Furthermore note that an irreducible
non-singular curve exists in a class $xb+yf$ exactly if the class lies in the
mentioned cone or is the element $b=(1,0)$ of negative self-intersection or 
one of the elements $f=(0,1)$ or $kb_{+}$ (with $k>0$ and $b_+=b_-+nf=(1,n)$ 
of $b_+^2=+n$) on the boundary of the mentioned cone; all of these classes
together with their positive linear combinations span the effective cone 
($x,y\geq 0$). 
Note that $c_1=2b+(n+2)f$ is positive only for $F_0$ and $F_1$ whereas for 
$F_2$ it lies only on the boundary of the positive cone.

For $F_2$ the Kodaira vanishing theorem 
tells us that $h^1(B,{\cal O}(C))=h^1(B,{\cal O}(K-C))=0$ 
if $-K+C=(x+2,y+4)$ is ample, i.e. for $y>2x, x>-2$; so clearly 
the superabundance will vanish for all ample $C=(x,y)$ (where even $x>0$)
and for $f$. Note that for $f$ we see directly that 
${\rm def}_{F_n}f=1=e(f)/2 + f^2$ (the expression from \deor\ )
and similarly for $b$ with $n\neq 0$ that ${\rm def}_{F_n}b=0$ and 
$e(b)/2 + b^2=1-n\leq 0$; so for the case $B=F_2$ the formal expression which
should give the number of actual deformations has to be 
modified, and - as this case is in a further respect somewhat exceptional
as we will see in the next subsection - we will {\it exclude} it.
Finally for $kb_+$ one sees that
$kb_+\cdot c_1(F_2)=4k>0$ making, according to the argument given above,
$s$ again vanish in this case although
$c_1(F_2)$ is not numerically effective in general.

\subsec{The deformations of $C$ in the elliptic surface ${\cal E}$}

{}From the Kodaira formula for the canonical bundle of ${\cal E}$ one has
an expression for $K_{{\cal E}}$ as a pull-back class (of course also
$F=\pi_{{\cal E}}^*p$ with $p\in C$)
\eqn\El{K_{{\cal E}}=\pi_{{\cal E}}^*K_C+\chi({\cal E},{\cal O}_{{\cal E}})F}
Then $\chi({\cal E},{\cal O}_{\cal E})$ is via Noether evaluated as
${1\over 12}e({\cal E})$ as $c_1({\cal E})$ has,
being a certain number of fibers, vanishing self-intersection. 
But $e({\cal E})=c_1(B_2)\cdot C$ as the elliptic fibration of the
Calabi-Yau has the discriminant $12c_1(B_2)$ so that
\eqn\chiE{\chi({\cal E},{\cal O}_{\cal E})=c_1(B_2)\cdot C}
Alternatively one can see from adjunction 
\eqn\adju{c({\cal E})=
c(C){{(1+r)(1+r+2c_1(B_2))(1+r+3c_1(B_2))}\over {1+3r+6c_1(B_2)}}}
that
\eqn\ok{\eqalign{c_1({\cal E})&= (e(C)-c_1(B_2)\cdot C)F \cr
e({\cal E})&= 12 c_1(B_2)\cdot C}}
Note that the number $d:=c_1(B)\cdot C$ has the following important 
interpretation:
from \El,\chiE {}       
one has $C_{{\cal E}}\cdot c_1({\cal E})=e_C-c_1(B)\cdot C_B$
(as $c_1({\cal E})$ is a pull-back class, i.e. a number of fibers) 
so that with adjunction $-e_C=C^2_{{\cal E}}-C_{{\cal E}}\cdot c_1({\cal E})$ 
inside ${\cal E}$ one gets 
\eqn\bquadr{C^2_{{\cal E}}=-c_1(B)\cdot C_B}
This gives after \scond\ another important criterion
\eqn\defE{C\cdot c_1(B_2)>0 \Rightarrow def_{{\cal E}}C_{{\cal E}}=0}
So except for the case $B=F_2, C=b=b_-$ we have no 
further deformations in the vertical direction; in the mentioned 
exceptional case one had ${\cal E}=b\times F$ showing the obvious deformation.
Note that in this case we had from the expression \deor\ that 
$e(b)/2+b^2=1-2=-1$. As in this exceptional case one has one deformation
in $Z$ and no deformation in the base $B$ which the $F$-theory side could see
directly we will {\it exclude} it from our final arguments.

\bigskip
\noindent{\it Examples}

To give some examples note that the first three cases $d=0,1,2$
leading to $e({\cal E})=0,12,24$ correspond to $b\times F, dP_9, K3$.
They occur in well-known circumstances if we choose $B_2=F_n$ with $n=2,1,0$. 
In that case occurs the
mentioned ${\cal E}$ over the base $C=b=b_-$ of self-intersection $-n$ 
of the Hirzebruch surface: the first case gives actually one of
the two exceptional divisors (the other is the base $F_2$ itself)
of the $STU$ Calabi-Yau ${\bf P}_{1,1,2,8,12}(24)$ which consists
in a ruled surface over the elliptic fiber (this is the just another
way to consider the product $b\times F$ fibered) \KV,\HKTY; 
the second case gives the $dP_9$ studied for example
in \KMV, which occurs not just in this set-up but 
over any exceptional curve ($b^2=-1, b\cdot c_1(B_2)=1$; rationality
and the second property imply the first) in a del Pezzo base; finally
for $F_0$ one gets the well-known $K3$ fibers, which occur of course
more generally also over each fiber of any Hirzebruch base (adjunction
shows $d=2$). 

\subsec{The deformations of $C$ in $Z$}

The total deformation space can be considered fibered together out of
the pieces investigated so far (cf. \DOd ).

Concerning $def_Z C$ consider 
\eqn\seqn{0\rightarrow N_{C/B}\rightarrow N_{C/Z}\rightarrow N_{B/Z}\otimes 
{\cal O}_C\rightarrow 0}
the last term being again $N_{C/{\cal E}}$.
To show that $H^0(N_{B/Z}\otimes {\cal O}_C)=0$ in order 
that $def_CB=def_CZ$ corresponds exactly with the already formulated condition
that there are no further deformations of $C$ in ${\cal E}$ 
if $C$ has there (!) negative self-intersection: for 
$deg N_{C/{\cal E}}=C^2_{{\cal E}}<0$ and $H^0$ vanishes just as in the
argument for $s=0$ given above.

\bigskip
\noindent{\it A remark about skew curves}

Although we will not treat this most general case (where one would have
to 'connect' blow-up's and three-branes on the $F$-theory side)
let us at least point
to the possibility that the total physical moduli space will include
even deformations of $C+a_fF$ away from the reducibility into base
(horizontal) and fiber (vertical) curves, leading to curves lying 'skew'
to the elliptic fibration. 

Until yet we have considered two possible cases where a heterotic 
five-brane wraps a curve either of class $a_fF$, i.e. of the 
purely vertical type - a reducible sum of $a_f$ fibers, 
or $W_B$, the purely horizontal type wrapping a base curve $C$. 
Of course that choice of a base curve realizing 
the cohomology class $W_B$ involves already considerable freedom in
itself; to give an example, for $B_2={\bf P}^2$ and $W_B=2L$ one
can choose either a smooth conic, giving a rational curve $C$ of
multiplicity $1$, or two crossing lines or even two times the same 
line, giving a rational curve $C$ of multiplicity $2$ (cf. also \RajDia). 
What concerns the distribution of the vertical class $a_f F$ in a
set of curves realizing it, one has considerable lesser freedom
as one can only make the in general disjoint $a_f$ fibers coinciding 
in some subsets corresponding to a decomposition of the number $a_f$.
So the set-up up to now would lead one in general to consider the wrapping 
of the reducible {\it curve} $C+a_fF$. Of course this realizes only a 
sublocus of very degenerate possibilities for a curve of 
{\it cohomology class} $C+a_fF$ as it may be actually possible 
to find an {\it irreducible} curve (of this cohomology class) 
inside $Z$, or more precisely, as we can assume, 
actually in the elliptic surface ${\cal E}$ over $C$ (cf. also \DOd ). 

This is by no means trivial as for example in the case
that the base curve $C$ is  rational and that
$h^{2,0}({\cal E})=0$ resp. $1$, i.e. ${\cal E}$ is either a 
rational elliptic ($dP_9$) surface or a $K3$. In that case
the class $D:=C+a_fF$ is not representable by an 
irreducible smooth curve because, although it is a section of the elliptic 
fibration of ${\cal E}$ and so rational, adjunction shows that 
$-2=D^2_{{\cal E}}-D\cdot c_1({\cal E})$; but in the $dP_9$ case 
$c_1({\cal E})=F$ and so the right hand side is $-1+2a_f-1$, i.e. $a_f=0$, 
whereas for $K3$ it is just $-2+2a_f$ showing again $a_f=0$. 
Here we used the fact that
$C^2_{{\cal E}}=-1,2$ on $dP_9,K3$ as one sees again from adjunction.

Note as a final remark that one has quite a remarkable {\it curve counting}. 
On $dP_9$ one has for the number $n_{1,\beta}$ of rational curves in the class
$C+\beta F$, as counted by the Gromov-Witten invariants and evaluated by mirror
symmetry \KMV, ($q=e^{2\pi i \tau}$; $\theta_{E_8}(\tau)$, the theta function
of the $E_8$ lattice being given by 
${1\over 2}\sum_{k=even}\theta^8_{k}(\tau)=E_4(\tau)$ 
with the Jacobi theta functions $\theta_k$ and the Eisenstein series
of weight $4$; $\eta$ is the Dedekind eta-function)
\eqn\ndp{\sum_{\beta} n_{1,\beta}q^{\beta}
=q^{{1\over 2}}{\theta_{E_8}(\tau)\over \eta^{12}(\tau)}}
Corresponding results for $\alpha C+\beta F$ with $\alpha > 1$ are given 
in \MNVW. Similarly in the $K3$ fiber ${\bf P}_{1,1,4,6}(12)$ (of the 
$STU$ Calabi-Yau of degree $24$) occurring over fibers of $F_2$ one has
\eqn\nstu{n_{\alpha ,\beta}=c_{STU}(\alpha \beta)}
where
\eqn\nstueisen{\sum_{n\ge -1}c_{STU}(n)q^n={E_4E_6\over \eta^{24}}(\tau)}
Note that these invariants can be really number of curves or more
generally Euler numbers of the deformation space of (or of the moduli space 
of flat $U(1)$ bundles over) the curve (in general one has to consider 
a certain obstruction bundle). In particular they don't tell us whether 
there is really an irreducible member besides that curve realization of the 
class $\alpha C+\beta F$ which is just given by the reducible curve 
consisting (for $\alpha =1$, say) in the base curve $C$ 
and a number of $\beta$ fibers. Further they 
don't tell us whether really a correspondingly big deformation space is 
realized in the elliptic surface ${\cal E}$, let alone in the Calabi-Yau
three-fold; even if an irreducible skew realization exists this does not 
necessarily mean that this realization is connected (in the sense of going 
smoothly through supersymmetric realizations) to the reducible realization.

\newsec{The singular case III: computation}

By now we are prepared with the necessary background, in order to handle 
the $\eta$-shift and see how the relations for the Hodge numbers of
the Calabi-Yau four-fold in terms of the heterotic data will change. 
To keep things as clear as possible we will separate the changes 
on the heterotic side from those appearing in the Calabi-Yau four-fold. 

\bigskip
\noindent{\it Heterotic Side}
\smallskip

On the route of our second strategy we found up to now two important things:
the shift in the index and the expression for the 
deformation (the unspecified $c_1$ refers to $B$)
\eqn\deltahetind{\eqalign{\Delta I&= -Cc_1 \cr
def_ZC&= {{C^2+Cc_1}\over{2}}}}
(Concerning the first relation note that this is the change in the
{\it functional form} of $I$ as a function of $\eta_1$, i.e. for the argument
to be given we are not interested in the change coming from just applying
the usual index expression to the new, shifted $\eta_1$. Concerning the
second relation note that, as indicated earlier, 
we exclude the exceptional case $B=F_2, C=b_-$.)

\bigskip
\noindent{\it F-Theory Side}
\smallskip

On the {\it F-theory side} the 
heterotic $\eta$-shift is understood as 
blowing up of the corresponding curve $C$ in the base of the Calabi-Yau 
four-fold which leads to a new divisor and so to a new Kaehler class, thus
\eqn\hee{h^{1,1}(X)\rightarrow h^{1,1}(X)+1}
also, this blow up leads to 
\eqn\gwb{g_{C}={{C^2-C\cdot c_1}\over{2}}+1}
new 3-cycles in $h^{2,1}(B_3)$ and also in $h^{2,1}(X)$, so
\eqn\hze{h^{2,1}(X)\rightarrow h^{2,1}(X)+g_C.}
{}Finally the change in $h^{3,1}(X)$ accounts for both effects, the 
index-shift $\Delta I=-Cc_1$ (as the index of the bundle moduli
maps to the deformations of the fourfold, cf. \hdg\ )
and the occurrence of the $C$-deformations
(which become geometrical on the $F$-theory side)
\eqn\hde{h^{3,1}(X)\rightarrow h^{3,1}(X)-Cc_1+def_BC}
{}From the relation \threebr\ for the number of three-branes 
we learn that the three changes in the Hodge numbers cancel out since 
(cf. \deor )
\eqn\zer{1-g_{C}-Cc_1+def_BC=0}
Thus, by expressing 
everything in data of the common base $B_2$, just as it
was done in \AC, we find that the desired matching $n_3=a_f$ of the number of 
heterotic five-branes wrapping elliptic fibers with the 
$F$-theory three-branes still holds in the more general situation with a 
non-trivial $W_B$, as the $\eta$-shift reduces this completely to the old
argument in the simpler situation of $W_B=0$.

Let us finally make a remark on the exceptional case.
For this let us compute the prediction for $h^{3,1}(X)$ 
from the heterotic side in case we
have an $E_8\times E_8$ bundle leaving no unbroken gauge group 
(and so having a smooth $X$ on the $F$ side) for the case of 
$B_2=F_2$ and wrapping the five-brane on the base $b$ of $F_2$.  
Let us assume that $t=0$ which means we have $B_3={\bf P}^1\times F_2$
on the $F$-side. Since $t=0$ we can either perform the $W_B$ shift in the
first $E_8$ or in the second one. The index computation
gives for the unshifted $E_8$ bundle $I=1336$ and for the shifted bundle
$1216$ therefore one has $h^{3,1}(X)=h^{2,1}(Z)+I+1+def_BC=243+1336+1216+1-1
=2795$. Now an independent computation (using toric geometry) of $h^{3,1}(X)$
\klem\ matches precisely the heterotic prediction. 

Note that we have used here the 'formal' expression for $def_BC$ which gives
$-1$ (so to speak an 'obstruction') whereas the actual number of deformations
of $b$ is $0$; further in $Z$ the number is even $+1$. 
This shows that this last heterotic deformation in ${\cal E}$ is, as one
expects, not directly visible on the $F$-theory side and that
on the other hand, concerning the deformations in the common visible $B=B_2$,
the $F$-theory side 'sees' the 'obstruction' too, as the $-1$ is reflected 
there.

\bigskip
We would like to thank A. Klemm, D. Morrison and J. Wahl for discussions. 

\listrefs
\bye